\documentclass[11pt,sort&compress]{elsarticle}
\usepackage[paper=a4paper,top=32mm, bottom=32mm, left=29mm, right=29mm]{geometry}
\makeatletter
\def\ps@pprintTitle{%
 \let\@oddhead\@empty
 \let\@evenhead\@empty
 \def\@oddfoot{\centerline{\thepage}}%
 \let\@evenfoot\@oddfoot}
\makeatother
\usepackage[T1]{fontenc}
\usepackage{amsmath}
\usepackage{graphicx}
\usepackage{amssymb}
\usepackage{mathrsfs}
\usepackage{marvosym}
\usepackage{booktabs}
\usepackage[english]{babel}
\usepackage{amsmath,amssymb,amsbsy,amstext, amsthm, simplewick}
\usepackage{graphicx}
\usepackage[framemethod=tikz]{mdframed}
\usepackage{amsfonts}
\usepackage{xcolor}
\usepackage{lipsum}
\usepackage{mathrsfs}
\usepackage{amsmath}
\usepackage{framed}
\usepackage{amssymb}
\usepackage{colortbl}
\usepackage{xcolor}
\usepackage{datetime2} 
\definecolor{green2}{cmyk}{0, 1, 0.5, 0}
\definecolor{lightgreen}{cmyk}{0.2, 0, 0.2, 0.2}
\definecolor{lightgray}{cmyk}{0.1,0.2,0,0.1}
\definecolor{lightgray2}{cmyk}{0.4,0.4,0,0.8}
\definecolor{black}{cmyk}{1.0,1.0,1.0,1.0}
\definecolor{brown}{rgb}{0.79, 0.16, 0.16}
\definecolor{cerulean}{rgb}{0.0, 0.48, 0.65}
\definecolor{blue2}{rgb}{0.0, 0.2, 0.4}
\definecolor{pc1}{rgb}{0.19, 0.0, 0.91}
\definecolor{royalazure}{rgb}{0.4, 0.2, 0.92}
\definecolor{darkcerulean}{rgb}{0.03, 0.27, 0.89}
\definecolor{magenta1}{rgb}{0.89, 0.08, 0.48}

\usepackage[subdued,defaultmathsizes]{mathastext}
\MTnonlettersobeymathxx  
\MTexplicitbracesobeymathxx 
\MTfamily {\ttdefault} 
\Mathastext [typewriter]
\usepackage{tikz}
\usetikzlibrary{tikzmark,decorations.pathreplacing,calc}

\RequirePackage{color}
\RequirePackage[colorlinks=true
,urlcolor=darkcerulean
,anchorcolor=black
,citecolor=cerulean
,filecolor=navyblue
,linkcolor=cerulean
,menucolor=navyblue
,pagecolor=navyblue
,linktocpage=true
,pdfproducer=medialab
]{hyperref}
\def\o2{{\sf O}(2)}
\def\u1{{\sf U}(1)}

\makeatletter
    \AtBeginDocument{\def\@citecolor{cerulean}}
    \AtBeginDocument{\def\@linkcolor{magenta}}
    \AtBeginDocument{\def\@filecolor{navyblue}}
     \AtBeginDocument{\def\@pagecolor{red}}
     \AtBeginDocument{\def\@urlcolor{pc1}}    
\makeatother

\usepackage[font={normalsize,sf}]{caption}

\newcommand{\floor}[1]{\left\lfloor #1 \right\rfloor}
\newcommand{\ceil}[1]{\left\lceil #1 \right\rceil}

\begin{document}
\begin{abstract} 
We consider the two-dimensional classical XY model on a 
square lattice in the thermodynamic limit using tensor 
renormalization group and precisely determine the critical 
temperature corresponding to the Berezinskii-Kosterlitz-Thouless 
(BKT) phase transition to be 0.89290(5) which is an improvement
compared to earlier studies using tensor network methods. 
\end{abstract}
\title{Critical analysis of two-dimensional classical XY model}
\author{Raghav G. Jha}
\address{Perimeter Institute for Theoretical Physics
\\31 Caroline Street North\\ Waterloo, 
ON, Canada\\{\vspace{0.5em}\normalfont }
}
\maketitle
{\vspace{-1.2em}
{\tableofcontents}
\begin{center}{\noindent\rule{\linewidth}{0.4pt}}\end{center}
\section{Introduction}
The XY model, also known as planar rotor model
is one of the most widely studied and the simplest two-dimensional spin 
model\footnote{The history of this model is not well established
but it appears that it was first discussed by Nambu 
in 1950 \cite{1950PThPh...5....1N} and was named XY for the 
first time by Lieb et al. \cite{Lieb:1961fr}}
with continuous symmetry because of the interesting 
features it possesses and the possibility of analytical treatment
using some reasonable approximations. 
The model has \u1 or \o2 symmetry corresponding to the rotation on a 
two-dimensional plane. The spins on each lattice site
interact via a nearest neighbour Hamiltonian.  
The model has numerous practical applications in studies related to 
superfluid helium, thin-films, superconductivity, liquid crystals, 
melting of two-dimensional crystals, and dielectric plasma transition from a dielectric phase 
(charges bound as neutral dipoles) to the conducting phase (free charges). 

The seminal work by Berezinskii and subsequently followed by
Kosterlitz and Thouless elaborated on the features 
of this model and provided an understanding of the 
infinite order phase transition without any order parameter. 
This phase transition does not accompany any symmetry breaking
and from the point of view of the 
Ginzburg-Landau description of phase transitions, this is a very special case. 
The change in the functional form of the correlation functions 
across the critical temperature signals the phase transition. 
The correlation length obeys an exponential dependence 
above the phase transition unlike the familiar power-law and it is infinite below 
the critical temperature. Any determination of the critical temperature by probing the 
system close to the critical temperature (i.e. $T_{\text{BKT}}$) is a challenging 
numerical problem. For temperatures below this critical value, 
there is a line of fixed points down to zero temperature. 
This is a special property that it shares for example
with supersymmetric $\mathcal{N}=4$ Yang--Mills in four dimensions. 
The XY model at zero temperature in the absence of an external field 
has all spins aligned in a particular direction corresponding to a 
ferromagnetic phase and shows long-range order
while at high temperatures it is in a 
symmetric paramagnetic phase in which all the 
directions on the plane are equivalent for all spins.

For many years, it was common folklore that two-dimensional 
nearest-neighbour spin systems cannot have 
phase transitions but the possibility of a finite temperature 
phase transition inferred by extrapolation 
from the high-temperature expansion 
was pointed out by Stanley and Kaplan in \cite{PhysRevLett.17.913}.  
This led to interesting developments eventually 
culminating in a series of works 
\cite{Berezinsky:1970fr,Kosterlitz:1973xp,Kosterlitz:1974sm}
after which many groups continued studying this model in great detail. 
Even though the two-dimensional XY model can be understood to a great extent
through analytical arguments based on mean-field theory, 
renormalization group methods, and by maps to 
other models in the same universality class 
(such as Coulomb gas, SOS (solid-on-solid) model),
the exact determination of the critical temperature on some fixed lattice
remains much a numerical problem. The study of the critical behaviour of the 
two-dimensional XY model has now been pursued for over four decades 
using high-temperature expansions (HTE), Monte-Carlo (MC) 
numerical approach, Hamiltonian strong coupling expansion and 
most recently using efficient tensor network (TN) algorithms. 
However, there is no general agreement on the critical temperature 
corresponding to the BKT transition and the available results
differ from one other by several error bars. 

In the majority of these explorations, 
MC approach (single-flip Metropolis) was applied. 
However, this computational approach
is prone to critical slowing down around the critical 
temperature and makes it highly inefficient. 
Several improved algorithms such as cluster algorithms\footnote{
For example, these include the Swendsen-Wang, Wolff,  and embedded 
cluster algorithms} where a group of spins is updated at once were then
formulated and was then used 
with parallel computations to locate the phase transition
on large systems close to the thermodynamic limit. 
Though until a few decades ago, 
the most preferred tool to study lower-dimensional ($d \le 2$) 
statistical systems were large-scale MC simulations, this is now 
giving way to tensor network techniques with two 
such studies already existing in the 
literature \cite{Yu:2013sbi, Vanderstraeten:2019frg}
for the model we consider in this work.  
The reason of this paradigm shift is clear -- in the 
last decade many efficient real-space
tensor network algorithms have been developed which have enabled 
studies of the critical and off-critical statistical systems in the 
thermodynamic limit by carefully exploiting the interesting region 
of the otherwise huge Hilbert Space. In the tensor network description, 
the physical properties of the system 
is encoded in the local tensors and by performing the coarse-graining 
sufficiently many times, we can understand the global properties in a 
more intuitive way.  Tensor networks 
have also been used to study spin-foam models in quantum 
gravity in various dimensions. They are 
also starting to play a crucial role in studying real-time path
integrals which are otherwise impossible to study using 
MC because of highly oscillatory form 
of integrals for which the importance sampling method fails. 
The tensor networks have also been useful in 
understanding some aspects of holography and understanding the 
connections between the many-body systems and gravity. 
For a review on tensor networks and its various applications, 
see \cite{Orus:2018}. It is now also becoming evident 
that these methods can also be used to study 
gauge theories and some works along this direction 
are \cite{Banuls:2019rao, Kuhn:2015zqa, Kadoh:2018hqq, Bazavov:2019qih, Franco-Rubio:2019nne}
\footnote{It is impossible to list the extended literature 
along these lines and we apologize in advance for all omissions.}. 

In this paper, we take a step in this direction by 
applying a real-space coarse-graining algorithm
and systematically determining the critical temperature for the 
topological transition in the XY model. 
By keeping a bigger fraction of the local space than achieved before, 
we have considerably improved on the earlier works. This resulted in our numerical estimate
as being the most precise achieved to date using any tensor network approach.
It also agrees with the most precise available MC result \cite{komura2012largescale}.  
We introduce a small external magnetic field that breaks 
the \o2 symmetry to determine the critical temperatures at
small field strengths by looking at the response of the network 
to the field by computing the magnetization and 
susceptibility.  We now briefly outline the structure of the paper. In Sec. \ref{sec1}, 
we discuss the phase transition and lack of ordering in two dimensions 
and explain the tensor construction of the model before discussing
the observables we study. In Sec. \ref{sec2}, we present the 
results for the free energy, magnetization, and determine the 
critical temperature from the peak of the susceptibility. 
We put down some numerical details in the Appendix
for the convenience of the reader. 

\section{\label{sec1}Phase transition, Tensor construction, and Observables} 

In two dimensions, a no-go theorem due to 
Mermin-Wagner-Hohenberg-Coleman (MWHC) states that it is 
not possible to have an ordered phase (long-range order)
\footnote{We mean the absence of long-range order in the 
sense explained by Polyakov in 
\cite{Polyakov:1987ez}. If one considers a large but finite
system and fixes the value of $\vec{S}$
at the boundary to be 1, the average value of  $\langle \vec{S} \rangle$
inside the system vanishes as system size goes to infinity. This
is equivalent to the fact that the angular deviation between spins 
increases with $r$ i.e. $\langle (\theta(r) - \theta(0))^{2} \rangle \sim \log(r)$.}
which implies that there is no possibility
of a phase transition from a disordered to an ordered phase 
accompanied by spontaneous symmetry breaking at 
any finite temperature. In high energy theory language, 
this is usually expressed as - there are no Goldstone bosons 
in two dimensions. The absence (or destruction) of long-range order is
due to the strong low-energy fluctuations. Since there is no 
long-range order, there is no true order parameter. 
However, there does exist a well-known phase transition 
where in the absence of an order parameter, the task of classification
of the phases depends on the behaviour of the correlation 
function. In the low-temperature phase of the XY model, 
there is a quasi-long-range order (QLRO) 
phase and we have: 
\begin{equation}
\langle\textbf{S}(\textbf{r})\cdot\textbf{S}(0)\rangle \sim  e^{-\eta \ln r} = \frac{1}{r^\eta}, 
\end{equation}
while in the high-temperature disordered phase, we have:
\begin{equation}
\langle\textbf{S}(\textbf{r})\cdot\textbf{S}(0)\rangle \sim  e^{-r/\xi},
\end{equation}
where \textbf{S}(\textbf{r}) is a planar spin of magnitude one on site 
$\textbf{r} = (\mathtt{x},\mathtt{y})$ 
of a square lattice, $\xi$ is the correlation length
and the exponent $\eta = T/(2\pi J)
$\footnote{The was first noted down by Peierls and Landau in 1935 and 1937 respectively}
is proportional to the temperature and is equal to 
its maximum value of 1/4 at $T_{\text{BKT}}$. 
This power-law behaviour is not just at critical point but 
for all temperatures less than  $T_{\text{BKT}}$.
The QLRO is destroyed by 
unbinding of the vortex and anti-vortex pair 
at some critical temperature $T_{\text{BKT}}$, 
which is now well-known as the Berezinskii-Kosterlitz-Thouless 
(BKT) transition. This transition has a topological character
associated with the proliferation of vortices. 
The BKT transition is a very special case of bypassing the 
MWHC theorem which rests on the fact that for
temperatures $T < T_{\text{BKT}}$, 
the algebraic (`power-law') decay of 
correlations functions means that 
there is \emph{no} long range order \emph{but} only QLRO 
where topological excitations/defects (vortex and anti-vortex with 
opposite topological charges) are bound together. 
At $T > T_{\text{BKT}}$, defects unbind and starts to grow rapidly  
and results in exponential decay (also referred to as ~`screening') 
given by: 
\begin{equation}
    \xi = 
    \begin{cases}
      \sim \exp\Big[\frac{\text{-C}}{\sqrt{\tau}}\Big], & T > T_{\text{BKT}}\\
      \infty , & T < T_{\text{BKT}}
    \end{cases}, 
  \end{equation}
where $\tau = (T - T_{\text{BKT}})/T_{\text{BKT}}$ and $C$ is a positive constant.
We now consider a square lattice with a spin on each lattice
site $\textbf{S}_{i}$. The Hamiltonian of the XY model is given by: 
\begin{equation}
\label{eq:Ham1} 
    \mathcal{H} =
    - J \sum_{\langle ij \rangle}\cos(\theta_{i}-\theta_{j})-h\sum_{i}\cos\theta_{i} ,
\end{equation}
where $\langle ij \rangle $ denotes the summation over the nearest 
neighbouring sites $i$ and $j$, $\theta_i$ is the spin angle at site $i$, and
we have set the modulus $| \textbf{S}_{i}| = 1$ with 
$J$ being the exchange coupling between the nearest
neighbour spins,
and $h$ the applied external magnetic field. We will set $J=1$
in what follows while the Boltzmann's constant, $k_{B}$, has already been set to 1.  
The Hamiltonian (\ref{eq:Ham1}) with $h=0$ has global \u1  symmetry: $\theta_{i} \to \theta_{i} +c$ 
which amounts to changing the angle by same amount at each lattice site 
and also has periodicity i.e. $\theta_{i} \to \theta_{i} + 2\pi n_{i}$.
We can write the partition function of the the model as:
\begin{equation}
 Z = \frac{1}{2\pi} \int \prod_i d\theta_i ~ \exp \Bigg(\beta \sum_{\langle ij \rangle}\cos 
 (\theta_i-\theta_j) + \beta h\sum_{i }\cos \theta_i \Bigg),
\end{equation}
where $\beta = 1/T$ is the reciprocal of temperature and angles vary 
continuously in the interval $[0,2\pi)$\footnote{The 
model where this interval is split in Q-intervals is known as
the Q-state Potts model which for $Q \to \infty$ is the XY model}. 
In order to construct the tensor representation, we 
decompose the Boltzmann factor (for $h=0$) in a basis 
(also known as Jacobi-Anger expansion) 
as follows:
\begin{equation}
\label{eq:expan 1} 
  \exp\Bigg({\beta\cos({\theta_i-\theta_j})}\Bigg)= I_{0}(\beta) + 
  \sum^{\infty}_{\nu=-\infty, \neq 0}{I_{\nu}(\beta)e^{i \nu (\theta_i-\theta_j)}},
\end{equation}
where $I_{\nu}(\beta)$ is the modified Bessel function of the first kind.
The partition function can then be written as \cite{Liu:2013nsa}:
\begin{equation}
  Z=\int \prod_i d\theta_i  \prod_{\nu_{ij},\mu_i} I_{\nu_{ij}} (\beta) 
  I_{\mu_i}(\beta h) e^{i\nu_{ij} (\theta_i-\theta_j) + i\mu_i\theta_i}  .
\end{equation}
By performing integration over the physical angular 
degrees of freedom $\theta_i$, 
we proceed to define a tensor (dual) as:
\begin{equation}\label{eqT}
\mathtt{T_{ijkl}} = \mathtt{T_{i,j,k,l}} = \sqrt{I_i(\beta)I_j(\beta)I_k(\beta)I_l(\beta)}
\overbrace{I_{i+k-j-l}(\beta h),}^{\text{for $h$=0, this enforces~} \delta^{i+k}_{j+l}}
\end{equation}
where indices $(i,j,k,l)$ denote the four legs of the 
tensor\footnote{We always use the order: left, right, up, down to denote a 
rank-four tensor in this work} and run from $ -\infty, \cdots, \infty$. 
The size of each leg, which 
is infinite in principle from the expansion formula
in (\ref{eq:expan 1}) is truncated down to run only in range
$ [\ceil{-\chi/2}, \cdots, \floor{\chi/2}]$, where we refer to $\chi$ as bond 
dimension. The reason this truncation is reasonable is that the series 
expansion coefficient $I_\nu(\beta)$ decreases quickly with increasing $\nu$
for all $\beta$ considered in this work.
Thus, we can truncate the series and approximate $\mathtt{T_{ijkl}}$ by a tensor with 
finite bond dimension with relatively high precision. However, the errors due 
to this truncation can depend strongly on the temperature and the 
external field $h$ as we will see later. This leads to a 
finite-dimensional tensor representation for the partition function:
\begin{equation}
Z = \mathrm{\bf{tTr}} \Big( \prod \mathtt{T_{ijkl}} \Big), 
\end{equation}
where $\textbf{tTr}$ denotes the tensor trace and product is over all sites.
\begin{figure}
\centering 
\includegraphics[width=0.81\textwidth]{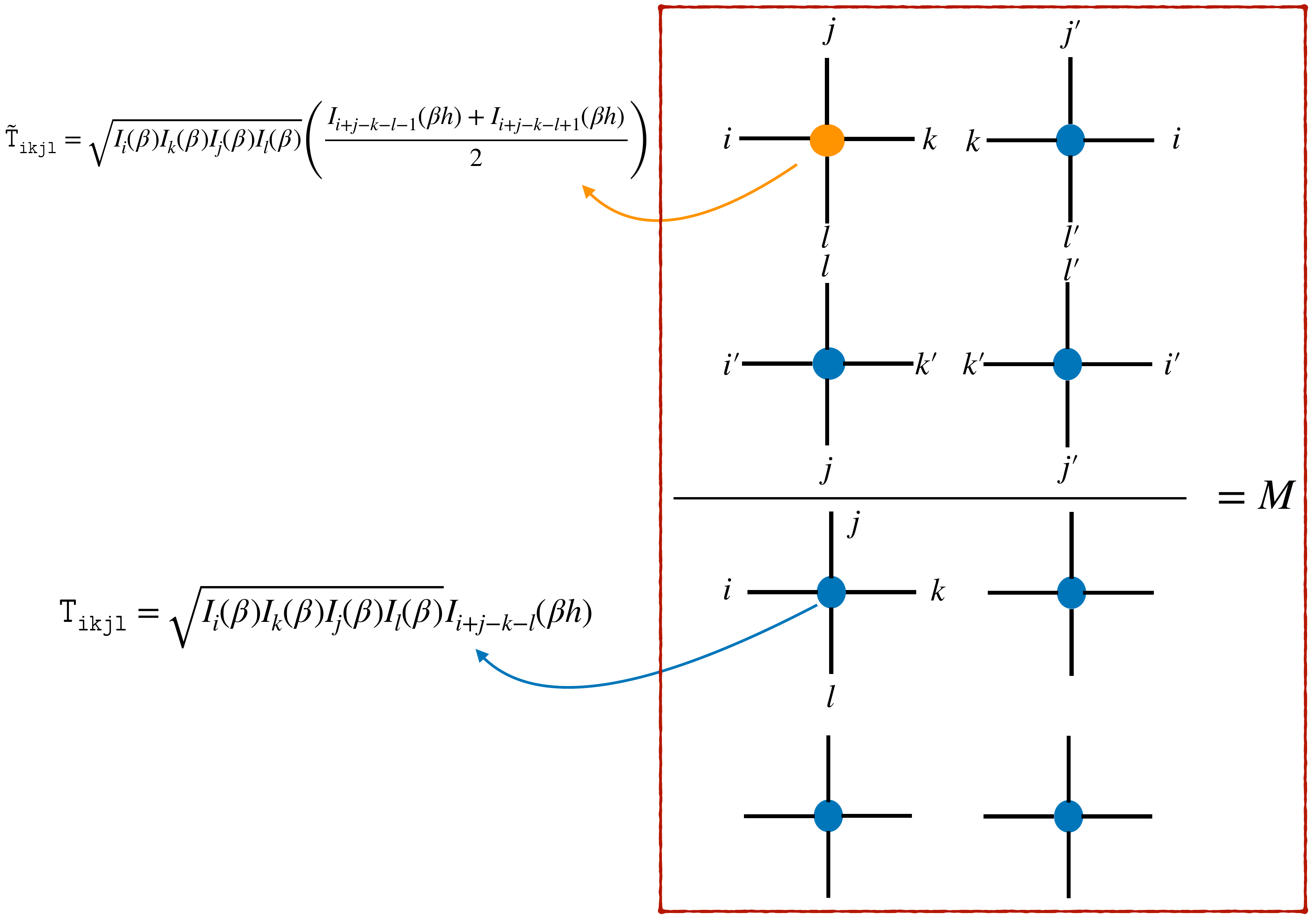}
\caption{\label{fig:mag1}The `observable' (sometimes also 
called ~`impure') tensor shown in orange is 
inserted in place of one of the usual (blue) tensors. We then evaluate 
magnetization by dividing the contraction of the network in the numerator and the  
denominator. The contraction is over the same index label and the ordering
and label of indices in the denominator is the same as in the numerator.} 
\end{figure}
It is well-known that in absence of magnetic field, 
the magnetization i.e. $M = \langle \vec{S_{i}^{x}}\rangle = \cos (\theta(\textbf{r}))$ is 
zero in $d=2$ for all non-zero temperatures. The average magnetization in 
presence of external field is evaluated as:
\begin{equation}
\label{eq:mag1} 
M = \frac{-\partial F}{\partial h} = \frac{1}{\beta} \frac{\partial \ln Z}{\partial h} = \textbf{tTr}\Bigg(
\sqrt{I_i(\beta)I_j(\beta)I_k(\beta)I_l(\beta)} \frac{I_{i+k-j-l-1}(\beta h) +  I_{i+k-j-l+1}(\beta h)}{2}\Bigg).
\end{equation}
Since the system has translational invariance, the average 
magnetization is the same for all spins. 
In principle, one can evaluate the magnetization 
by taking the numerical derivative of the logarithm of the partition 
function with respect to external field as in (\ref{eq:mag1}), but, 
this method is prone to errors and is not suited for 
determining the critical temperature. We evaluate the 
magnetization by inserting an additional tensor in the tensor network as
described in detail in Figure \ref{fig:mag1}.
The response function of the system when the external field is turned on
is given by magnetic susceptibility defined as:
\begin{equation}
\mathbb{S} = \frac{\partial M}{\partial h}\Bigg \vert_{T}.
\end{equation}. 
\section{\label{sec2}Results \textit{\&} Conclusion} 
We use higher-order tensor 
renormalization group (HOTRG)
\cite{Xie2012CG} using periodic 
boundary conditions to study this model. This real-space 
coarse-graining scheme has been very useful in
studying quantum and classical statistical systems in
various dimensions \cite{Unmuth-Yockey:2018xak, Akiyama:2019xzy}.
Instead of coarse-graining along a constant time-slice and then constructing 
the transfer matrix, we have used simultaneous coarse-graining along 
$\texttt{x}$ and $\texttt{y}$ direction. We then use the singular value
decomposition (SVD) to truncate the tensor back to the original 
size while introducing a small error in the description of the system. 
This truncation becomes severe with decreasing $h$. 
\begin{figure}
\centering 
\includegraphics[width=0.81\textwidth]{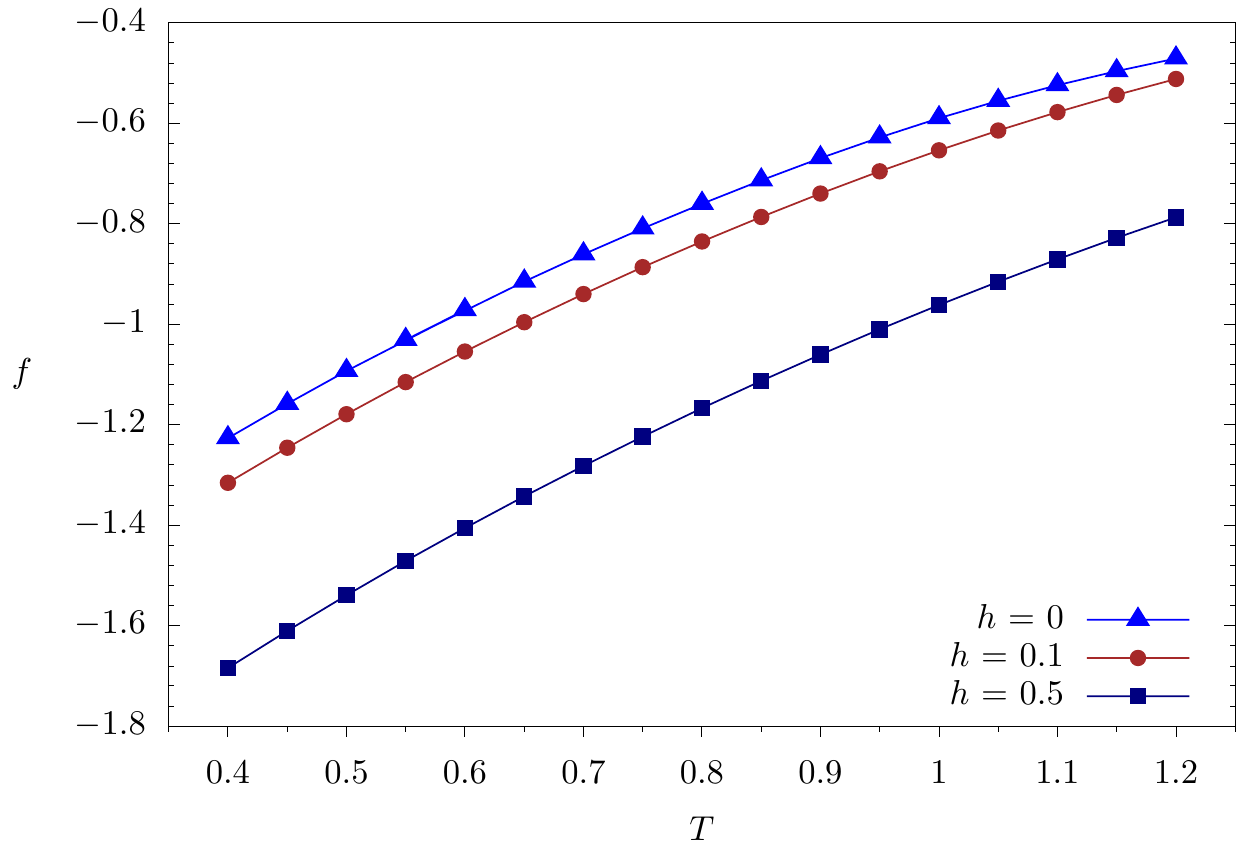}
\caption{\label{fig:free}Free energy against temperature for different values of magnetic field.
The system size is $2^8 \times 2^8$ and $\chi=25$.}
\end{figure}
We first calculate the free energy (using \ref{eq:free1}) and show 
the its variation with temperature for 
different external magnetic field $h$ in Figure \ref{fig:free}. Unlike in the Monte Carlo
simulations, the free energy (or density) is the simplest quantity to measure 
in the tensor network formulation. We checked that the free energy 
at $T = 1.0$ is the same up to 
an error of $\mathtt{10^{-4}}$ whether 
($N_{\mathtt{x}}N_{\mathtt{y}},\chi$) is $\mathtt{(2^{60}, 47)}$ or 
$\mathtt{(2^{16}, 25)}$\footnote{$N_{\mathtt{x}}N_{\mathtt{y}}$ denotes the number of sites along the $\mathtt{x}$ and $\mathtt{y}$  direction
respectively. One of these is about 200 times faster than the other even 
without worrying about the memory}. One important feature of the phase transition in this system is that 
all derivatives of the free energy are zero as $ T \to T_{\text{critical}}$ and 
there is no way of locating the transition 
by looking at any of its derivatives since there are no 
discontinuities at the critical temperature. 
This makes this system an exception to the usual 
Landau-Ginzburg paradigm. They belong to the class of topological phase
transitions that are not characterized by a Landau order parameter. 
Our determination of the critical temperature was done
by looking at $\mathbb{S}$ for different $h$ and then 
taking the $\displaystyle \lim_{h \to 0}$ of the critical temperature using an ansatz
(see Appendix). In an external magnetic field $h$, the magnetization behaves like 
$M \sim h^{\frac{1}{\delta}}$ at the critical temperature 
and the critical exponent describing the response to the field i.e. $\delta$ is given by
$\delta = (d+2-\eta)/(d-2+\eta) = 15$.
\begin{figure}
\centering 
\includegraphics[width=0.81\textwidth]{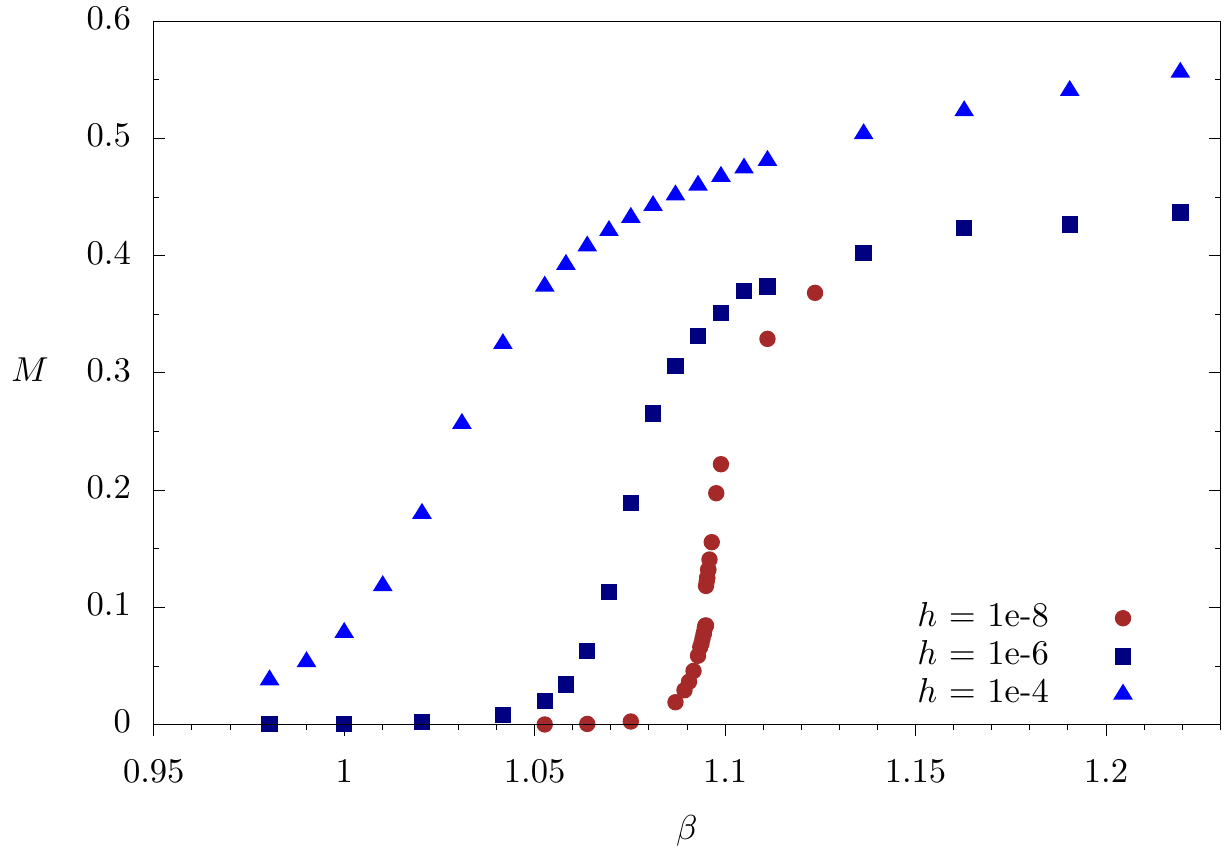}
\caption{\label{fig:magne1}The dependence of magnetization on
inverse temperature for different $h$.}
\end{figure}
We evaluate the magnetization for various
$h$ in the thermodynamic limit 
(lattice of size $\mathtt{2^{50} \times 2^{50}}$)
and show a small subset of those measurements in Figure \ref{fig:magne1}.
In order to determine the critical temperature, we
find out the temperature corresponding to the 
peak of susceptibility for different field strengths
as shown in Figure \ref{fig:sus}. 
We then extrapolate our numerical data and find 
$T_{\text{critical}} = \mathtt{0.89290(5)}$
in the limit of vanishing field shown in Figure \ref{fig:pplot}. 
In order to precisely determine the critical temperature, 
it was essential to go to very small values of the external field 
which in turn was strongly dependent on the choice of 
$\chi$. We initially started with $\mathtt{\chi = 41}$ and found that it 
was not possible to explore less than $h \approx \mathtt{10^{-7}}$ while 
accurately determining the peak of susceptibility. 
We believe it was fortuitous that this bond dimension ($\chi$) 
was sufficient to determine the critical temperature reasonably well and 
critical exponent to a within few percent of accuracy in \cite{Yu:2013sbi}.
For instance, even with $\mathtt{\chi = 53}$, we could not precisely 
determine the critical exponent, $\delta$, but certainly been able 
to reduce few orders of magnitude of error on the previous estimate of $T_{\text{critical}}$.   
As compared to earlier work, we 
kept almost double the number of states at each coarse-graining 
step because of using a slightly improved version of HOTRG
and memory available. These 
extra states were crucial to explore 
much smaller magnetic fields down to $h = \mathtt{10^{-15}}$ 
compared to approximately 
$h =\mathtt{10^{-7}}$ in \cite{Yu:2013sbi}. 
For instance, in \cite{Yu:2013sbi}, the critical temperature for 
$h = \mathtt{1.5 \times 10^{-7}}$ was found to be $T = \mathtt{0.9172}$ 
while we found $T = \mathtt{0.9215(3)}$. This is a clear sign that increasing $\chi$ on 
basically similar lattice size still helps the precision as one probes 
lower temperature and goes to zero field limit.
\begin{figure}
\centering 
\includegraphics[width=0.81\textwidth]{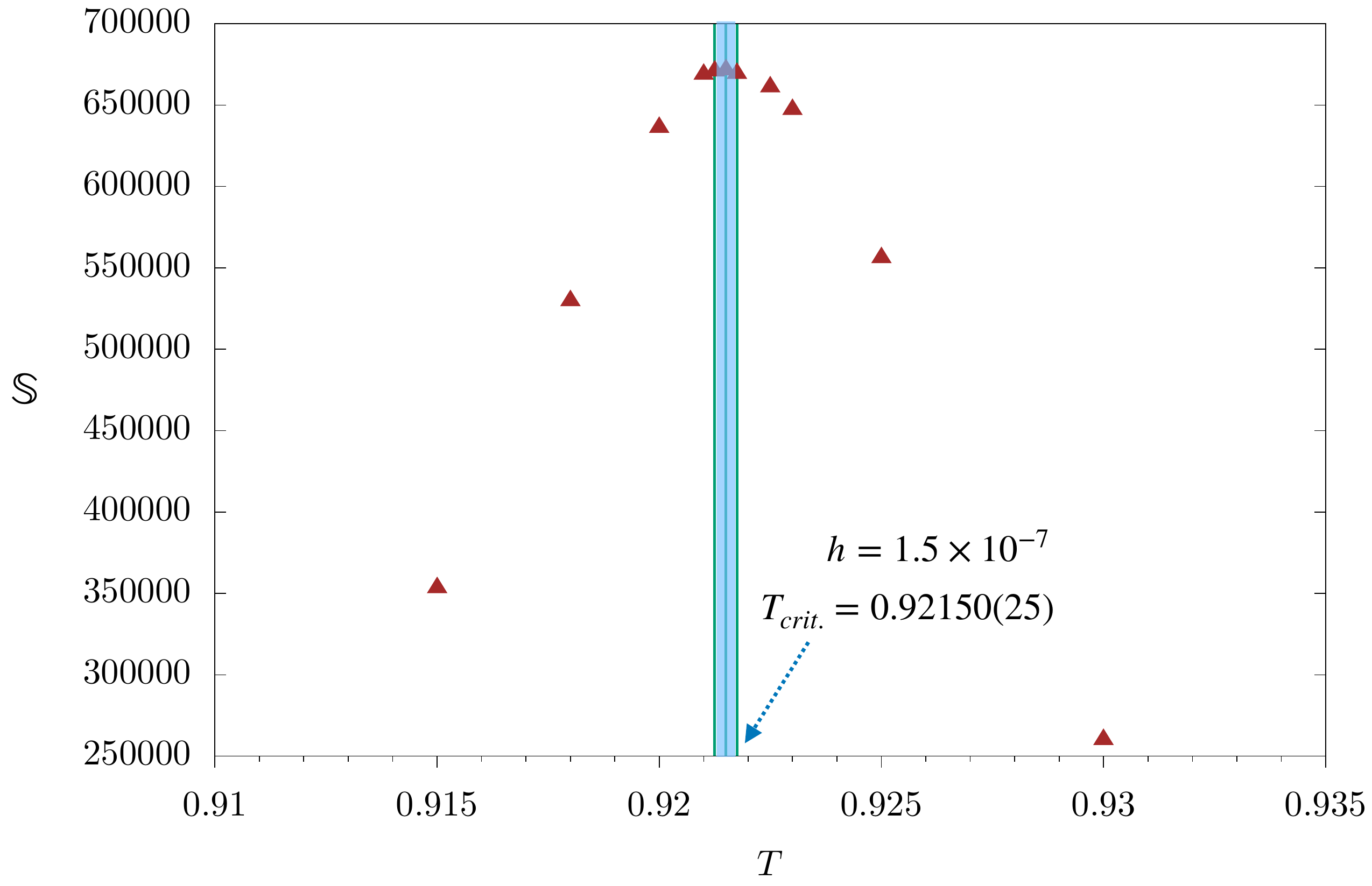}
\caption{\label{fig:sus}A representative example of the determination of the critical 
temperature for a given magnetic field. The vertical shaded band
denotes the error in the estimate of the critical temperature.}
\end{figure}
\begin{table}
\begin{center} 
  \renewcommand\arraystretch{1.4}  
  \addtolength{\tabcolsep}{7.15 pt}   
 \begin{tabular}{||c |c | c| c|} 
 \hline
  \textbf{\textsc{Method}} &  \textbf{\textsc{Year}} & \textbf{\textsc{System Size}}  & \textbf{$T_{\text{critical}}$} \\ [0.2ex] 
  \hline  \hline
  Monte Carlo  \cite{PhysRevB.45.2883} & \texttt{1992} & $\mathtt{2^{9} \times 2^{9}}$  & \texttt{0.89400(500)} \\ 
  \hline 
  HTE  \cite{PhysRevB.47.11969}  & \texttt{1993} &    --     & \texttt{0.89440(250)} \\ 
  \hline 
  Monte Carlo  \cite{PhysRevB.52.4526} & \texttt{1995} &  $\mathtt{2^{8} \times 2^{8}}$    & \texttt{0.89213(10)} \\ 
  \hline 
   Monte Carlo  \cite{Hasenbusch:2005xm}  & \texttt{2005} & $\mathtt{2^{11} \times 2^{11}}$    & \texttt{0.89294(8)} \\ 
\hline
  HTE  \cite{PhysRevE.79.011107}  & \texttt{2011} &    --     & \texttt{0.89286(8)} \\ 
\hline
  Monte Carlo  \cite{komura2012largescale}  & \texttt{2012} & $\mathtt{2^{16} \times 2^{16}}$    & \texttt{0.89289(5)} \\ 
  \hline
  Monte Carlo  \cite{hsieh2013finitesize}  & \texttt{2013} & $\mathtt{2^{9} \times 2^{9}}$    & \texttt{0.89350(10)} \\ 
  \hline 
 Higher-order TRG \cite{Yu:2013sbi}   & \texttt{2013} & $\mathtt{2^{40} \times 2^{40}}$    & \texttt{0.89210(190)}  \tikzmark{FooA} \\ 
 \hline 
Uniform MPS \cite{Vanderstraeten:2019frg} &  \texttt{2019} &  --   & \texttt{0.89300(10)}  \\ 
  \hline
  Higher-order TRG  [This work]  & \texttt{2020} & $\mathtt{2^{50} \times 2^{50}}$    &\texttt{0.89290(5)}  \tikzmark{FooB} \\ 
  \hline 
\end{tabular}
\begin{tikzpicture}[remember picture, overlay]
\draw [thick,decorate,decoration={brace}]
($(pic cs:FooA) + (0.74, 0.54)$)   --  ($(pic cs:FooB) + (0.95,-0.2)$) node [midway,xshift=0.45cm] {\texttt{TN}};
\end{tikzpicture}
\caption{\label{tab1}The estimate of the temperature for the BKT transition in 
classical XY model with different methods. The three most recent estimates have all been done 
using real-space renormalization with tensor networks (TN).}
\end{center}
\end{table}

\begin{figure}
\centering 
\includegraphics[width=0.81\textwidth]{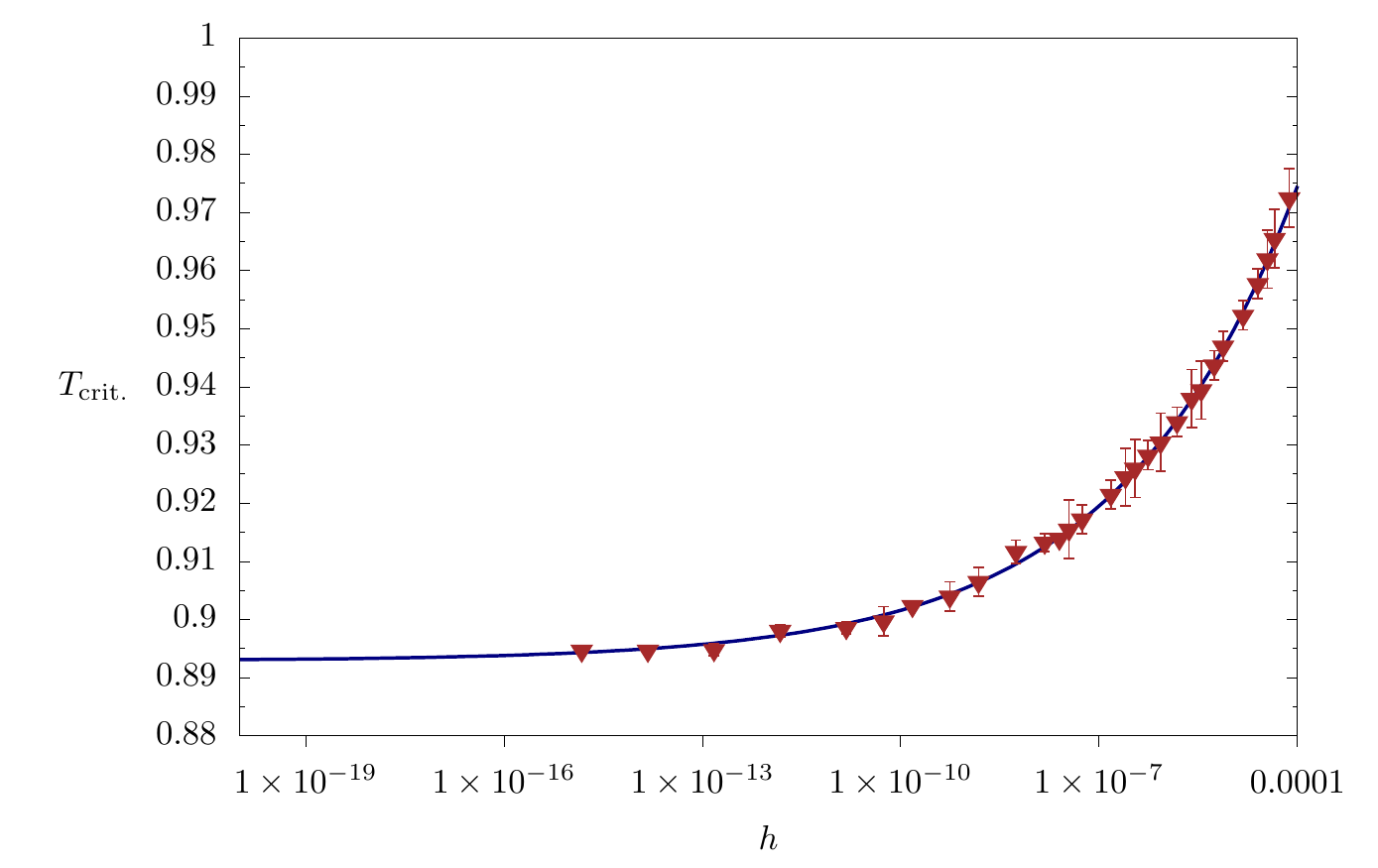}
\caption{\label{fig:pplot}The dependence of critical temperature on 
external magnetic field $h$. We fit the data to the ansatz and obtain 
$\displaystyle \lim_{h \to 0} T_{\text{crit.}}$ = 0.89290 (5).
The errors have been scaled by a factor of 10 for visibility on the plot
and determined as in Figure (\ref{fig:sus}). 
The data is for square lattice of size $2^{50} \times 2^{50}$
with $\chi_{\text{max.}} = 53$.}
\end{figure}

We now briefly mention some earlier works locating the critical 
temperature for this model (see Table \ref{tab1} for the list). 
Very recently, using uniform MPS and a well-defined method of extracting correlation length for 
discrete system as given in \cite{PhysRevX.8.041033}, this model was studied 
in \cite{Vanderstraeten:2019frg} and 
$T_{\text{critical}} = \texttt{0.8930(1)}$ was found. This was an improvement 
over the old work using tensor networks which found 
$T_{\text{critical}} = \texttt{0.8921(19)}$ \cite{Yu:2013sbi}. 
This model was also studied in \cite{komura2012largescale}
using 256 GPUs on a $65536 ~ (= 2^{16}) \times 65536$ lattice 
using Monte Carlo.
Some earlier studies differ significantly from our estimate and we 
guess that it might be because the lattice is too small. It was shown 
in \cite{Vanderstraeten:2019frg} that already around $T = 0.95$, 
the correlation length can be more than a thousand sites.
We note that among all these, the numerical study using 
Monte Carlo on the largest lattice is indeed also the most 
precise estimate to date which we have been able to match using 
tensor networks in this work. Though the procedure in this paper 
is similar in spirit to the one in \cite{Yu:2013sbi}, we have been 
able to refine the results in that paper by going to a large bond
dimension. For instance, for $h \sim 1.5 \times 10^{-4}$, 
the peak appears at $T = 0.98$ in the earlier work
while we found that the peak is at $T = 0.9810(5)$. 
A more prominent difference is observed for $h \sim 1.5 \times 10^{-7}$, 
where earlier study found $T = 0.9172$ while we 
deduced $T = 0.9215(3)$. 

In this work, the estimate of the critical temperature is the most precise 
that has yet been achieved using any tensor network approach and are level with 
precision from Monte Carlo methods. In the preliminary stages, 
we also implemented the tensor network renormalization (TNR) \cite{2015PhRvL.115r0405E} 
algorithm which removes the short-range entanglement by using 
disentanglers and makes the study at and near criticality more efficient 
compared to the other standard numerical RG approaches.
We found that while the results for free energy
and magnetization obtained using TNR 
are consistent with HOTRG, the complexity and 
the scaling of the TNR code was less encouraging. 
However, we believe that a systematic large-scale study 
of this model using efficient disentangling algorithms might be 
able to further improve on the 
errors we have reported here and extract the critical exponents
precisely but it would be challenging. In the future, it will be interesting to explore phase transitions in 
$\emph{frustated}$ XY model and other spin models such as 
J-Q model exhibiting continuous symmetry using these 
methods.

\subsection*{\textsc{Acknowledgements}}
\noindent We thank Simon Catterall, Judah Unmuth-Yockey, Yuzhi Liu, Haiyuan Zou, 
Suraj Shankar, Guifr\'{e} Vidal, Stefan K\"uhn, Glen Evenbly,  Nikhil Kalyanapuram, and 
Utkarsh Giri for discussions. The numerical computations 
were done on - Symmetry (Perimeter's HPC system).
Research at Perimeter Institute is supported in 
part by the Government of Canada through the Department of 
Innovation, Science and Economic Development Canada 
and by the Province of Ontario through the Ministry of 
Colleges and Universities.

\section*{\label{appen0}Appendix A: Numerical details}
In this appendix, we elaborate on some numerical details and 
steps to determine the critical temperature.
We normalize the tensor $\mathtt{T_{\nu}}$ at each coarse-graining step (denoted by $\nu$)
by the maximum element of the tensor having a total of $\chi^4$ elements 
which can be implemented in Python using 
NumPy library as $\mathtt{norm_{\nu} = numpy.max(T_{\nu})}$. 
We can then calculate the free energy density ($f$) 
from these normalization factors, see for instance 
\cite{zhao2015tensor}: 
\begin{equation}
\label{eq:free1}
f  =  -\frac{1}{\beta}\Bigg(\sum_{\nu=0} ^{N} 
\frac{\log(\mathtt{norm_{\nu})}}{4^\nu} + \frac{\log(Z_{N})}{4^N}\Bigg) \
= - \frac{1}{\beta 4^N} \Bigg(\sum_{\nu=0} ^{N} 
\log(\mathtt{norm_{\nu}}) 4^{N-\nu} + \log Z_{N} \Bigg), 
\end{equation}
where $Z_{N} = \textbf{tTr}(\mathtt{T_{N}})$ is calculated from the tensor contraction 
after the last step of CG and $4^{N}$ is the lattice volume and $N$ is the 
number of times we do coarse-graining along $\mathtt{x+y}$ direction.
To compute expectation values, one needs to 
insert the appropriate ~`observable' tensor
in the network. By inserting $\mathtt{\tilde{T}}$, 
which is just the derivative of $\mathtt{T}$ with respect to $h$,
we evaluated the magnetization for a
range of magnetic fields at various temperatures. 
The tensor, $\mathtt{\tilde{T}_{\nu}}$, is normalized by 
$\mathtt{norm_{\nu}}$ as well and is calculated as
$ M = \frac{\text{tTr}(\mathtt{\tilde{T}_{\nu}})}{\text{tTr}(\mathtt{T_{\nu}})}$ 
at each step. The diagram representing this contraction is
shown in Figure \ref{fig:mag1}. 
Though the free energy density converges rather quickly, we
really need to work in the thermodynamic limit with as large $\chi$
as possible to precisely evaluate magnetization.
In order to compute
the magnetization for $\mathtt{10^{-15}} \le h < \mathtt{10^{-9}}$, we
used single-precision float ($\texttt{np.float32}$) rather 
than the default double precision. 
This enabled us to go up to $\mathtt{\chi=53}$ compared to 
$\mathtt{\chi=47}$ for $h \ge \mathtt{10^{-9}}$. 
Once the magnetization $M$ is computed for different $h$, we evaluate 
$\Delta M / \Delta h \vert_{T, h_{\text{mid}}}$ and plot it against 
temperature and determine the critical temperature corresponding to the peak 
of the susceptibility. The data for the determination of critical temperature
for various $h$ can be found at \cite{Jha2:2020}.  
We used Python's $\texttt{\text{curve\_fit}}$ 
which uses Levenberg-Marquardt algorithm
and fit our data to the ansatz: 
$T_{c} = T_{c,h=0} + ah^{b}$, with $a$ and 
$b$ as two fitting parameters. The result of the fit is: \newline 
\[ T_{c,h=0}  = \mathtt{0.89290(5)}, 
\mathtt{a = 0.364(2)}, \text{and} ~ \mathtt{b = 0.162(1)} ,\] \newline 
and is shown by the solid line in Figure (\ref{fig:pplot}).

\vspace{10mm} 

\bibliographystyle{utphys}
\bibliography{v2}
\end{document}